\documentclass[twocollum]{aa}
\usepackage{graphicx}
\usepackage{nicefrac}
\usepackage{multicol}
\usepackage{natbib,twoopt}
\usepackage[draft,breaklinks=true]{hyperref} 
\usepackage[varg]{txfonts}
\usepackage{epsfig,times,lscape,rotating}
 \bibpunct{(}{)}{;}{a}{}{,}   
 \newcommandtwoopt{\citeads}[3][][]{\href{http://adsabs.harvard.edu/abs/#3}%
                                        {\citealp[#1][#2]{#3}}}
 \newcommandtwoopt{\citepads}[3][][]{\href{http://adsabs.harvard.edu/abs/#3}%
                                        {\citep[#1][#2]{#3}}}
 \newcommandtwoopt{\citetads}[3][][]{\href{http://adsabs.harvard.edu/abs/#3}%
                                        {\citet[#1][#2]{#3}}}
 \newcommandtwoopt{\citealtads}[3][][]{\href{http://adsabs.harvard.edu/abs/#3}%
                                        {\citealt[#1][#2]{#3}}}
 \newcommandtwoopt{\citeyearads}[3][][]%
   {\href{http://adsabs.harvard.edu/abs/#3}{\citeyear[#1][#2]{#3}}}
\begin{document}

\title{Gaps, rings, and non-axisymmetric structures in protoplanetary disks - Emission from large grains}

   \author{J.P. Ruge
          \inst{1}
          \and
          M. Flock\inst{2,3}
          \and
          S. Wolf\inst{1}
          \and
          N. Dzyurkevich\inst{4}
          \and 
          S. Fromang\inst{3}
          \and
          Th. Henning\inst{5}
          \and
          H. Klahr\inst{5}
          \and
          H. Meheut\inst{3,6}
          }

   \institute{Universit\"at zu Kiel, Institut für Theoretische Physik und Astrophysik,
Leibnitzstr. 15, 24098 Kiel, Germany \\
              \email{ruge@astrophysik.uni-kiel.de}
         \and
Jet Propulsion Laboratory, California Institute of Technology, Pasadena, California 91109, USA
\and
CEA UMR AIM Irfu, SAP, CEA-CNRS-Univ. Paris Diderot, Centre de Saclay,
F-91191 Gif-sur-Yvette, France         
          \and
          Laboratoire de radioastronomie, UMR 8112 du CNRS, \'{E}cole normale
sup\'{e}rieure et Observatoire de Paris, 24 rue Lhomond, F-75231 Paris
Cedex 05, France
          \and
          Max-Planck-Institut für Astronomie, K\"onigstuhl 17, 69117 Heidelberg, Germany   
          \and
          Laboratoire Lagrange, Universit\'e C\^ote d'Azur, Observatoire de la C\^ote d'Azur, 
          CNRS, Bd de l'Observatoire, CS 34229, 06304 Nice cedex 4, France\\
             }

   \date{}

\abstract{}{Dust grains with sizes around (sub)mm are expected to couple only weakly to the gas motion in regions beyond $10\, \rm au$ of circumstellar disks. In this work, we investigate the influence of the spatial distribution of such grains on the (sub)mm appearance of magnetized protoplanetary disks.}{We perform non-ideal global 3D magneto-hydrodynamic (MHD) stratified disk simulations including particles of different sizes ($50\, \rm \mu m$ to $1\, \rm  cm$), using a Lagrangian particle solver. Subsequently, we calculate the spatial dust temperature distribution, including the dynamically coupled submicron-sized dust grains, and derive ideal continuum re-emission maps of the disk through radiative transfer simulations. Finally, we investigate the feasibility to observe specific structures in the thermal re-emission maps with the Atacama Large Millimeter/submillimeter Array (ALMA).}{Depending on the level of the turbulence, the radial pressure gradient of the gas and the grain size, particles settle to the midplane and/or drift radially inward. The pressure bump close to the outer edge of the dead-zone leads to particle trapping in ring structures. More specifically, vortices in the disk concentrate the dust and create an inhomogeneous distribution of the solid material in the azimuthal direction. The large-scale disk perturbations are preserved in the (sub)mm re-emission maps. The observable structures are very similar to those expected to result from planet-disk interaction. Additionally, the larger dust particles increase the brightness contrast between gap and ring structures. We find that rings, gaps and the dust accumulation in the vortex could be traced with ALMA down to a scale of a few astronomical units in circumstellar disks located in nearby star-forming regions. Finally, we present a brief comparison of these structures with those recently found with ALMA in the young circumstellar disks of HL Tau and Oph IRS 48.}{}
\titlerunning{Emission from large grains}
\authorrunning{Ruge et~al.}
\maketitle

\section{Introduction}
\label{sec:intro}

   \begin{figure}
	\centering
       \resizebox{\hsize}{!}{\includegraphics{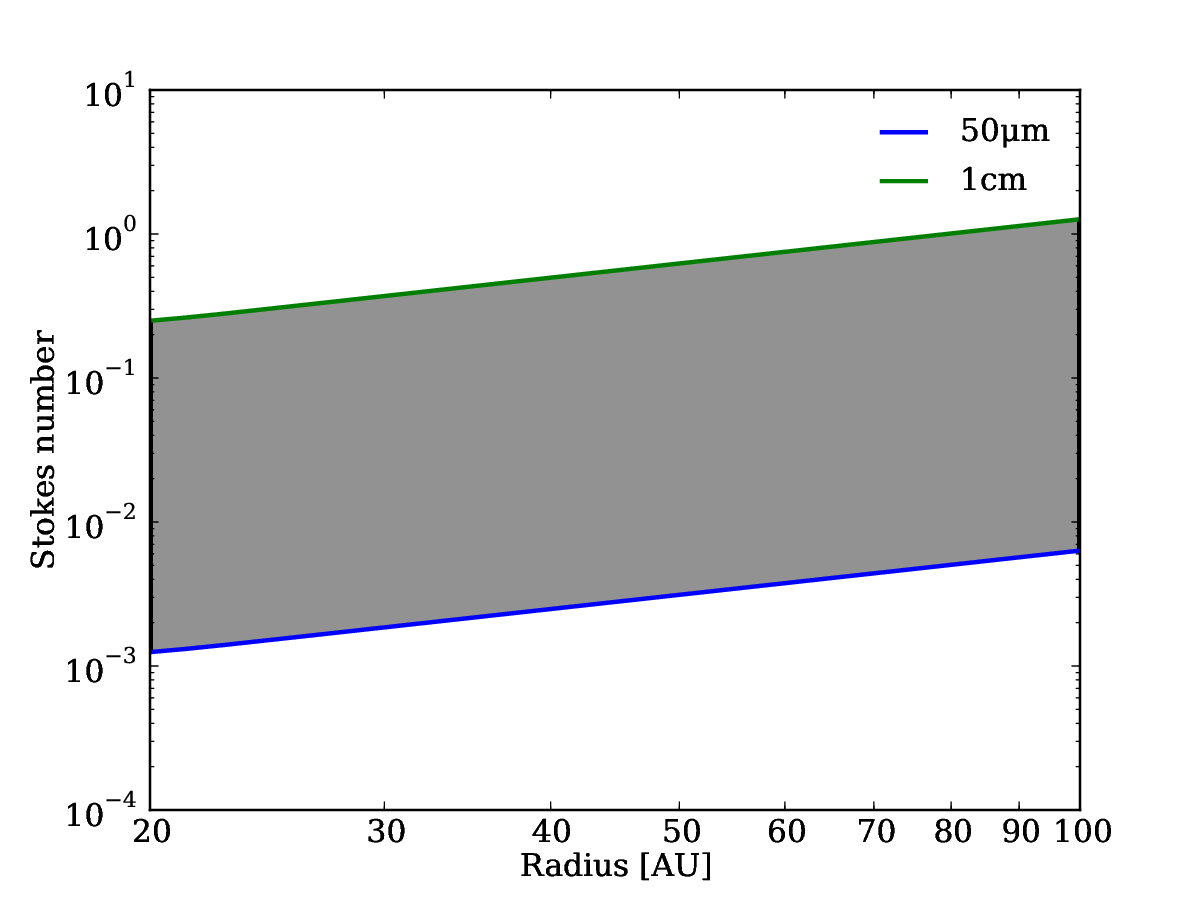}}
  \caption{Radial profile of the initial particles Stokes number at the midplane. The green and the blue line mark the regime of the considered particle sizes.}
  \label{fig:stokes}
   \end{figure}

Recent observations of protoplanetary disks have revealed the
existence of a wealth of structures with various properties, from dust
asymmetries and concentrations in the outer regions of protoplanetary
disks (e.g., Oph IRS 48: 
\citealtads{2013Sci...340.1199V}, LkH$\alpha$ 330:
\citealtads{2013ApJ...775...30I}, SAO 206462 and SR 21:
\citealtads{2014ApJ...783L..13P}, or in HD 142527:
\citealtads{2013Natur.493..191C,2013PASJ...65L..14F}) to the amazing
image of HL Tau (ESO press release 1436 and
\citealtads{2015arXiv150302649P}) obtained during ALMA science
verification phase using its most extended configuration and revealing
without ambiguity a multiple dust ring structure. These exciting
observations have often been interpreted as the signature of the
presence of a young planet in the disk. This is because the dust
spatial distribution is believed to trace the underlying gaseous
structure of the disk.

Indeed, while small dust grains of submicron size are well coupled to
the motion of the gas in circumstellar disks, this is not the same
for particles of (sub)mm size that form from collisional growth due to
various underlying physical processes
\citepads[e.g.,][]{2014prpl.conf..339T}. For these larger particles,
the coupling between the grains and the gas reduces
\citepads{2000prpl.conf..533B} and large--scale gas structures, like
radial pressure gradients or vortices are able to locally concentrate
these partially decoupled particles
\citepads{bar95,tan96,1997Icar..128..213K,cha00,joh04,2009A&A...493.1125L,2012MNRAS.419.1701R,2012A&A...545A.134M,2012A&A...538A.114P,2013A&A...550L...8B,
  2013ApJ...775...17L,zhu14}. Various mechanisms creating such large-scale
structures have been proposed, such as variations in the accretion
stress at the gas ionization boundaries
\citep{var06,2009A&A...497..869L,2010A&A...515A..70D,fau15,lyr15,Flock},
sublimation zones \citep{kre07,bra08,dzy13,zha15}, sintering zones \citep{oku15}, planet-disk interactions
\citep{2004A&A...425L...9P,2014ApJ...785..122Z}, photo-evaporation
(\citealtads{2012ApJ...747..103E}) or gravitational disk instabilities
\citep{ber01,vor05}. Further investigation confirmed their
observability, e.g. for planet-disk interactions
(\citealtads{2005ApJ...619.1114W}, \citealtads{2012A&A...547A..58G},
\citealtads{2013A&A...549A..97R}),  self gravitating disks
(\citealtads{2010MNRAS.407..181C}, \citealtads{2014MNRAS.444.1919D}, \citet{poh15}, \citet{tak16})
or vortices (\citealtads{2002ApJ...578L..79W,2012A&A...538A.114P}). As
we can see, the presence of a planet in the disk as the origin of the
recently observed structure is only one possibility among others and must
be examined critically. In addition, those disk systems are evolving on timescales of several millions of years with ongoing dust growth and reprocessing \citep{wil11,wol12} but the exact age of the observed system is difficult to estimate.

In this context, we have shown in \citetads{Flock} that it is possible 
to explain pronounced dust rings and local concentrations in magnetized disks 
without a planet. 
At the outer edge of the dead-zone, the response of the magneto-rotational instability (MRI) to variations of the gas density supports the formation of a large-scale gap and jump structure in the gas density distribution. The surface density maximum triggers the Rossby wave instability \citepads{1999ApJ...513..805L,les09,lin10,lin12a,lin12b}, which is a special form of Kelvin Helmholtz instability \citep{pap84}, resulting in the formation of a vortex. 
In the present study, we further investigate the spatial and temporal distribution of large dust particles (radii ranging from $50\, \rm \mu m$ to $1\, \rm  cm$) in such a disk model. We focus on the impact of these dust grains on the (sub)mm appearance of the circumstellar disk. 
For this purpose, we calculate the motion of each particle individually using a Lagrangian method and simulate the thermal re-emission maps based on the resulting dust density distribution with the continuum radiative transfer code MC3D (\citealtads{1999A&A...349..839W,2003CoPhC.150...99W}). The results are compared with our previous model that include only small dust grains \citepads{Flock}. 
Therefore, we make use of the same disk setup, 
which is characterized by a total disk mass of $0.042\, \rm M_\odot$, a disk radius ranging from $20\, \rm au$ to $100\, \rm au$, a dead-zone from $20\, \rm au$ to $40\, \rm au$ , a central star with an effective temperature of $4000\, \rm K$, a luminosity of $0.95\, \rm L_\odot$ , and a stellar mass of $0.5\, \rm M_\odot$.

\section{Modeling techniques}
\label{sec:hydro}

\begin{figure*}[!]
\begin{minipage}{18cm}
  \includegraphics[width=9cm]{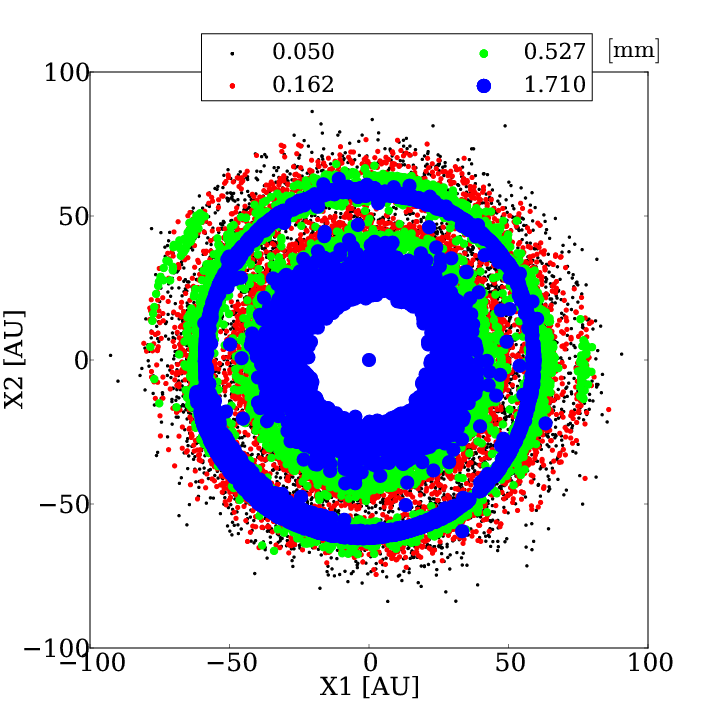}
  \includegraphics[width=9cm]{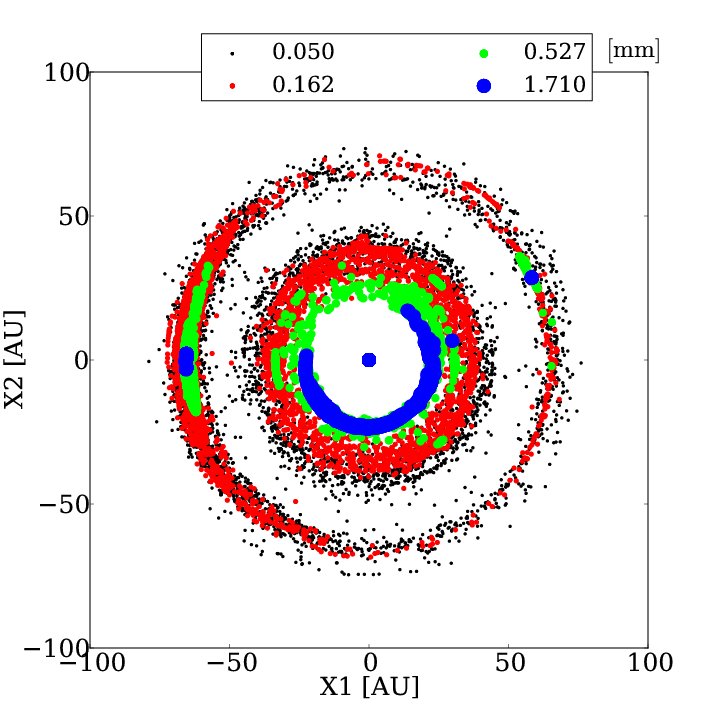}
\end{minipage}
\caption{Scatter plot of every tenth particle in four selected size
  bins of the disk in face-on orientation. Left: Time step after $150$
  inner orbits (corresponding to $18,\!900$ years of
  evolution). Right: Time step after $600$ inner orbits ($75,\!600$
  years).  The particle size is given in units of $\rm mm$. After
  $150$ inner orbits ($18,\!900$ years), the
  particles are concentrated into rings. After $600$ inner orbits
  ($75,\!600$ years), the vortex has broken the
  ring structures and has concentrated the larger particles.}
\label{fig:particles}
\end{figure*}

We use non-ideal global 3D MHD stratified simulations of a circumstellar disk model as described in our previous work (see model \texttt{D2G\_e-2} in Section 2 and Fig. 7 in \citealtads{Flock}). The dust particles with perfect coupling to the gas motion have sizes of $a_\text{min} = 0.005\, \rm \mu m \leq a \leq 0.25\, \rm \mu m = a_\text{max}$ and they are called small dust particles in the present work. The grains are homogeneous, compact and spherical, and follow the size distribution
\begin{equation}
 \cfrac{\text{d}n(a)}{\text{d}a} \propto {a}^{-3.5}.\label{eq:grainsizedist}
\end{equation}
In addition to the distribution of the small dust, we consider spherical particles with the following radii: $50 \, \rm \mu m$, $90 \, \rm \mu m$, $0.2\, \rm mm$, $0.3 \, \rm mm$, $0.5 \, \rm mm$, $0.9 \, \rm mm$, $1.7\, \rm mm$, $3.1\, \rm mm$, $5.5\, \rm mm$, and $10\, \rm mm$. We refer to them as large dust particles. We consider the external forces of gas drag and gravity.
To calculate the grain trajectories, we solve the equation of motion in the spherical coordinates system ($r$,$\vartheta$,$\phi$):
\begin{eqnarray}
\rm \frac{\partial v_r^{par}}{\partial t} &=& \rm \frac{(v_{r}^{\text{gas}} - v_r^{par})}{\tau_\text{t}} + \frac{l_{ \vartheta}^2}{r^3} + \frac{l_\phi^2}{r^3 (\sin{ \vartheta})^2 }-\frac{G M_\star}{r^2}\\
\rm \frac {\partial l_{\vartheta}}{\partial t} \, \, &=& \rm \frac{(v_{ \vartheta}^{\text{gas}}\cdot r - l_ \vartheta)}{\tau_\text{t}} + \frac{ l_\phi^2 \cdot cos{ \vartheta}}{r^2 (\sin{ \vartheta})^3}\\
\rm \frac{\partial l_{\phi}}{\partial t}\, \, &=& \rm \frac{v_\phi^{\text{gas}}\cdot r \cdot \sin{ \vartheta} - l_\phi}{\tau_\text{t}} \label{glg:positions}
\end{eqnarray}
with the coupling time $\tau_\text{t}$, the gravitational constant G, the stellar mass $\rm M_\star$ and the angular moments of the particle $\rm l_ \vartheta = v_ \vartheta^{\text{par}} \cdot r$ and $\rm l_\phi = v_\phi^{\text{par}} \cdot r \cdot \sin{ \vartheta}$. The particle time integration is done using a Leapfrog integrator. We use outflow boundary conditions for the gas and the particles. In this work, we focus on the Epstein regime, which is valid when the particle size a is smaller than
$\nicefrac{9}{4}$ of the mean free path of the gas molecules \citep{1977MNRAS.180...57W}. In this case, the coupling time $\tau_\text{t}$ can be expressed as:
\begin{equation}
\tau_\text{t} = \cfrac{\rho_\text{d} a}{\rho_\text{g} c_\text{s}},
\end{equation}
 with the solid density $\rm \rho_d=2.7\, g.cm^{-3}$, the gas density $\rm \rho_g$ and the sound speed $c_s$. The Stokes number $\rm St = \tau_t \cdot \Omega$ of the large dust particles is shown in Fig.~\ref{fig:stokes}. Initially, we place $50,\!000$ particles in each size bin, randomly distributed in the simulation domain. The initial velocity of the particles is set to match the local Keplerian velocity. In Appendix~\ref{ap:benchrt} we show, that the effect of the number of particles and a different distribution of the particles onto the results remains small. \par
   \begin{figure}[!]
	\centering
       \resizebox{\hsize}{!}{\includegraphics{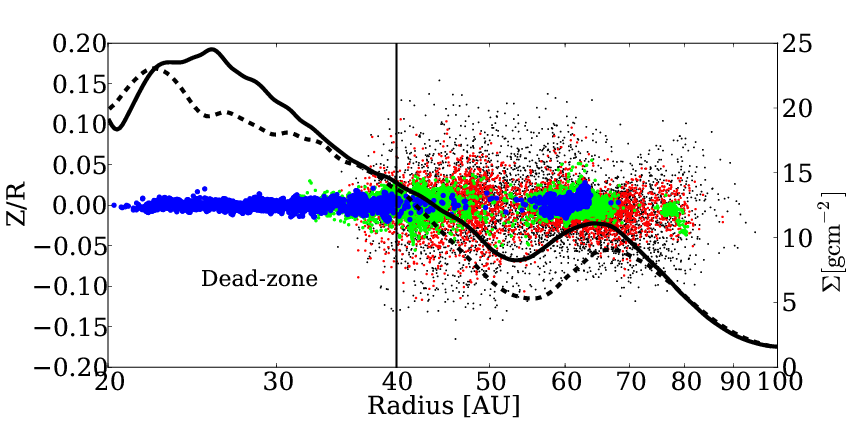}}
       \resizebox{\hsize}{!}{\includegraphics{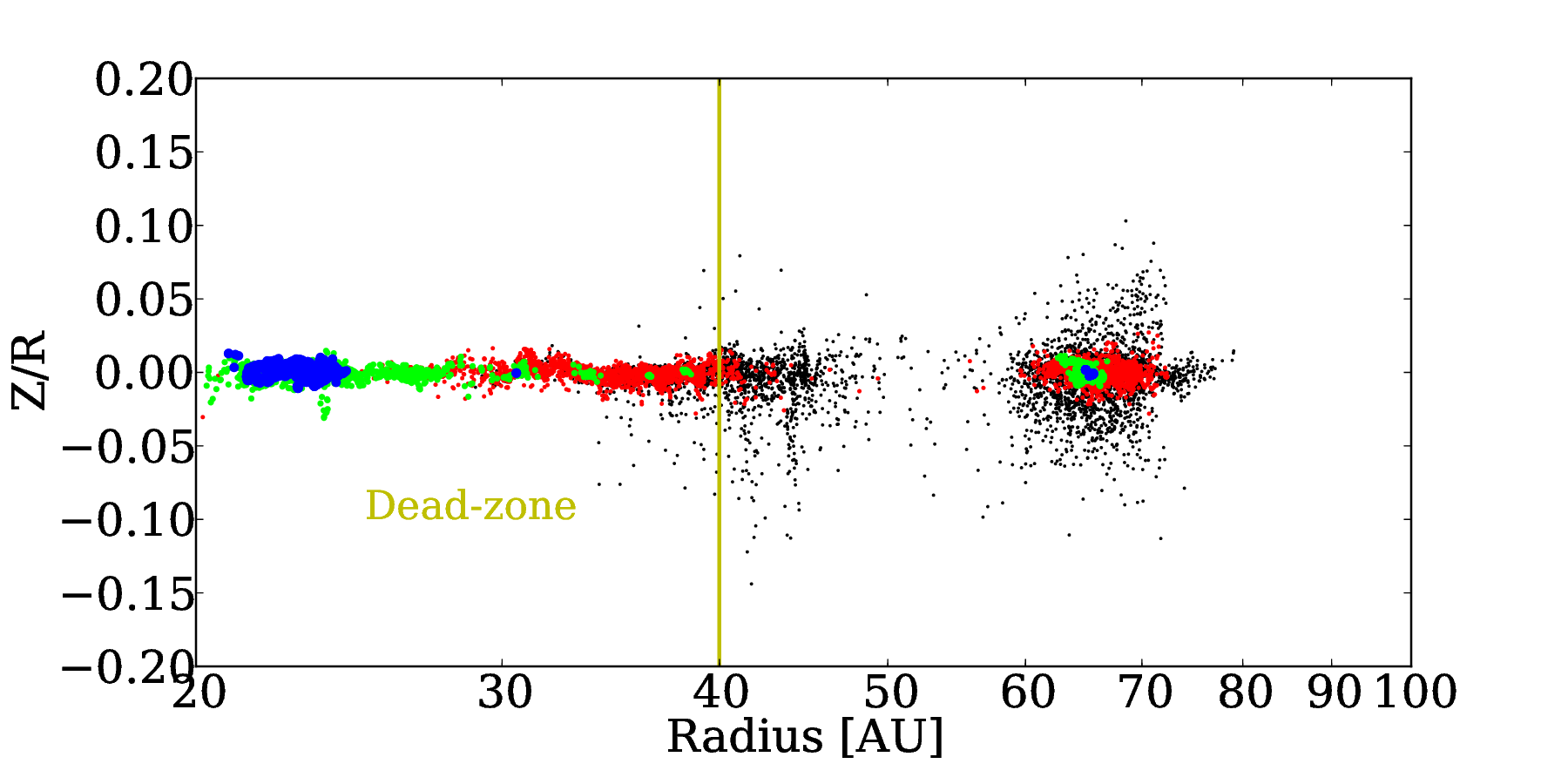}}
  \caption{Scatter plot of every tenth particle in four selected size bins of the disk in edge-on orientation for the time step after $150$ inner orbits ($18,\!900$ years) (top) and $600$ inner orbits (bottom). The dead-zone outer edge is marked in the plot. Particles are color--coded according to their size as in figure~\ref{fig:particles}. The surface density profile of the disk models is overplotted. The black solid line shows the surface density profile of the disk after $150$ inner orbits ($18,\!900$ years), the dashed line shows this profile for the disk after $600$ inner orbits ($75,\!600$ years).}
  \label{fig:sigma}
   \end{figure}
Two snapshots of the simulation will be considered in the following: one after 150 inner orbits of the disk, corresponding to $18,\!900$ years of disk evolution, and one after 600 inner orbits, corresponding to $75,\!600$ years. After 150 orbits of the inner disk, the (sub)mm dust particles have accumulated into ring structures (see Fig. \ref{fig:particles}, left) due to a pressure jump at the outer dead-zone edge (see Fig. \ref{fig:sigma}).  In Fig.~\ref{fig:sigma}, we project the distribution of the larger particles in the 2D R-$\rm Z/R$ plane. 
We remind that the larger particles are located along an azimuthal ring, while they slowly oscillate in the vertical direction.
Another time snapshot could show a different distribution for z > 0 and z < 0. During the following hundreds of orbits, the Rossby wave instability generates a vortex that concentrates dust particles depending on their
size. We plot the grains spatial distribution that results at 600
orbits on the right panel of Fig.~\ref{fig:particles}. In agreement
with the findings of \citetads{2012A&A...545A.134M}, we observe an
increased concentration of larger particles towards the vortex center. We
note that we can only observe the increase of the dust to gas mass
ratio and not its saturated value as we do not consider the dust
back-reaction onto the gas. Moreover, we do not resupply particles at
the outer domain boundary and the presented accumulation of solid
material in the computational domain should thus be seen as a lower limit. During the full simulation run, we observe an increase of the dust to gas mass ratio by 20\% when averaging over a disk scale height. However, we note that on scales smaller than a tenths of a disk scale height the increase of the dust component can be larger. Only future very high resolution dust and gas simulation including the back-reaction of the dust on the gas and a consistent resupply of particles can investigate the effect of the solid component on the gas.

Inside the dead-zone (for $r < 40\, \rm au$), all large particles remain at the midplane (see Fig. \ref{fig:sigma}). Outside of the dead-zone, the $50\, \rm \mu m$ particles are still efficiently mixed in the vertical direction up to about one gas scale height. In Fig. \ref{fig:sigma}, we overplot the surface density of the gas at the two selected timesteps. The gas surface density
of the gap and jump is decreasing between the two timesteps outward of
$40\, \rm au$ (see Fig. \ref{fig:sigma}). Such a decrease in gas
density leads to reduced coupling between the particles to the gas,
thereby favoring the maintenance of the ring on long timescales. The
dynamical process is conducive to the formation of a dust ring by the
evolution of the gas density and the subsequent decoupling of the dust
was recently demonstrated in a study by \citetads{2014arXiv1411.5366M}.

\begin{figure}[!]
	\centering
       \resizebox{\hsize}{!}{\includegraphics{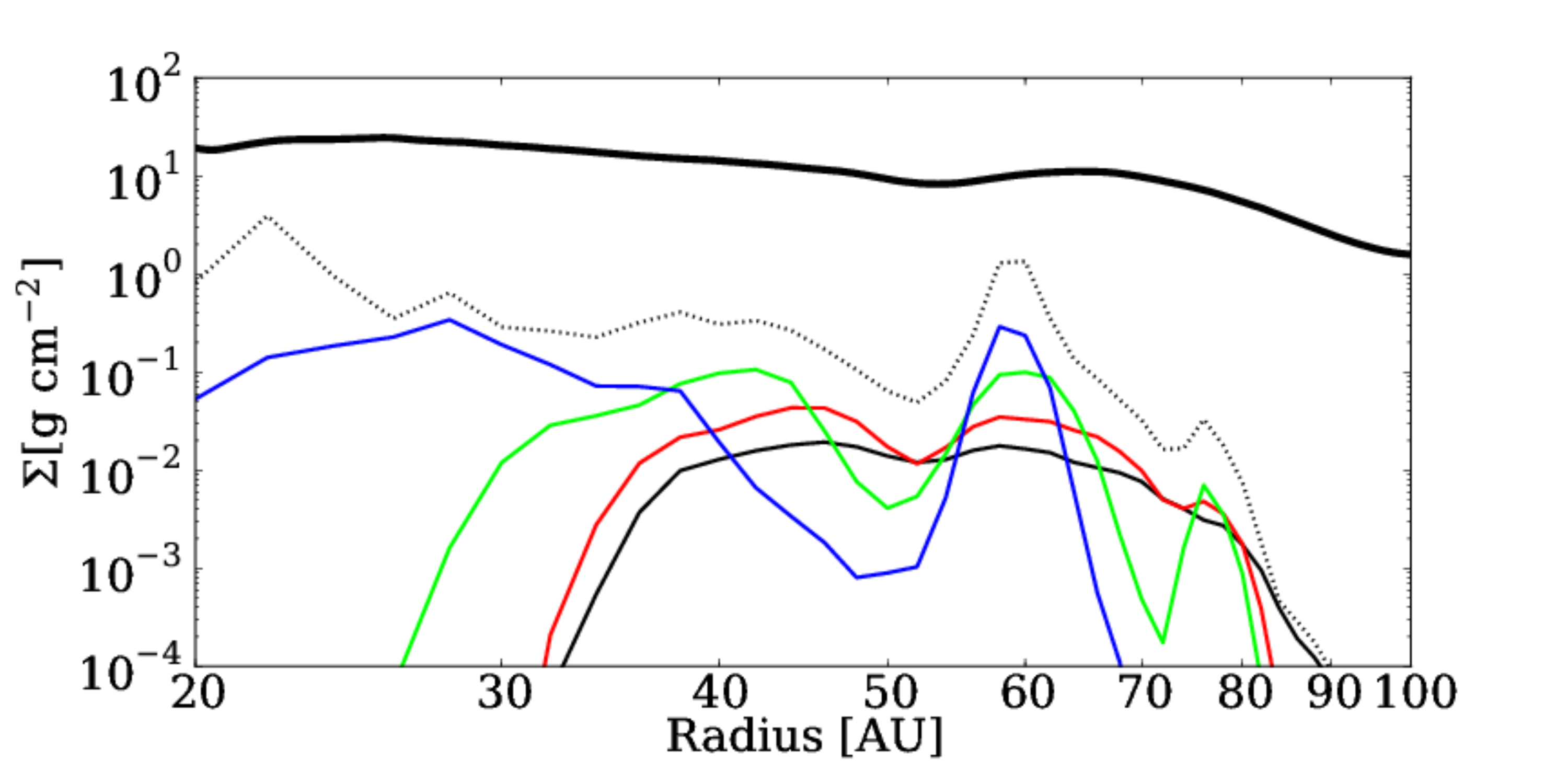}}
       \resizebox{\hsize}{!}{\includegraphics{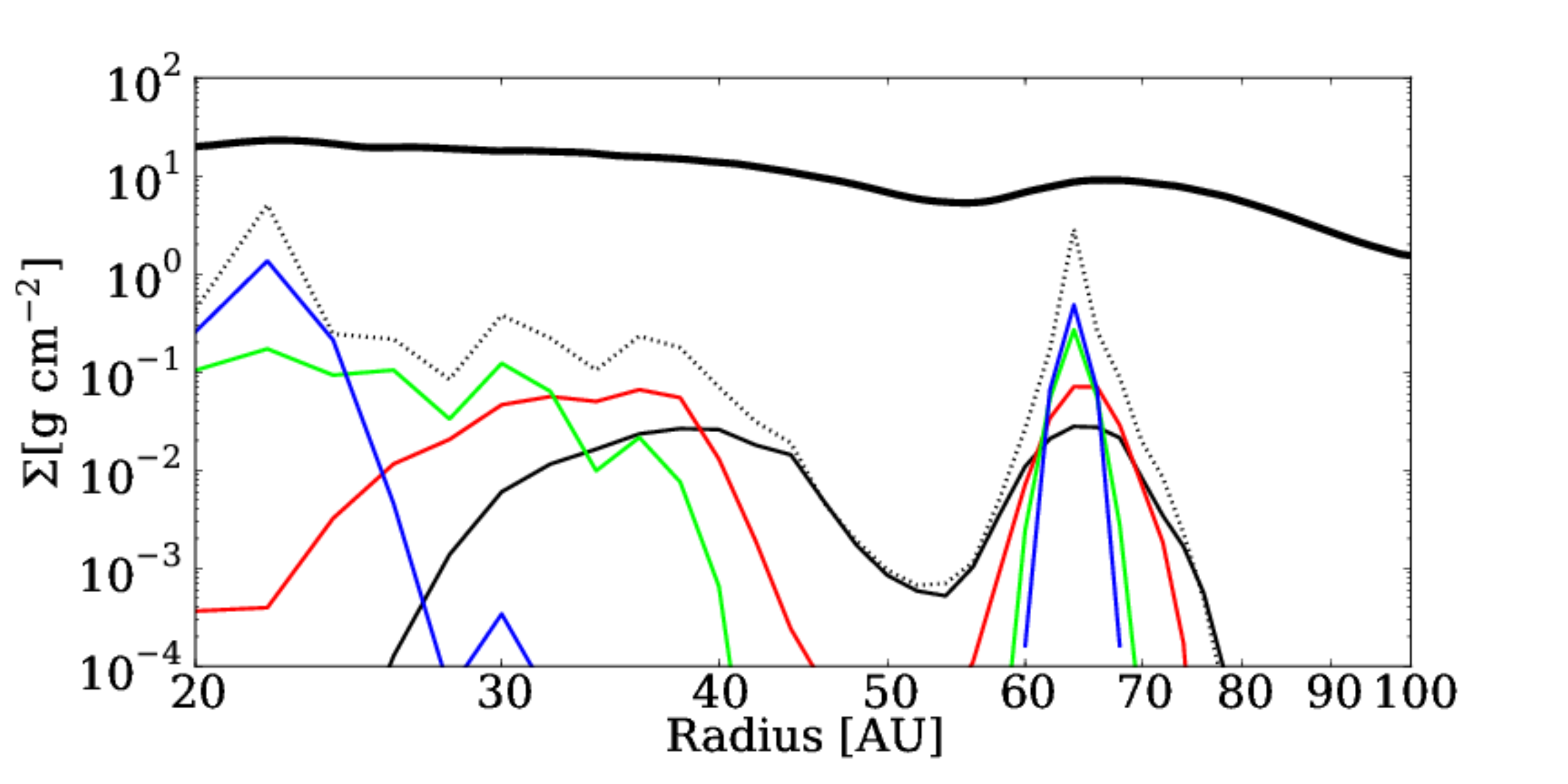}}
  \caption{Surface density plot for the gas (thick black solid line), the total dust (dotted line) and the 4 representative dust components, 1.71 mm (blue), 527 $\rm \mu$m (green), 162 $\rm \mu$m (red) and 50 $\rm \mu$m (black), for the time step after $150$ inner orbits ($18,\!900$ years) (top) and $600$ inner orbits (bottom).}
  \label{fig:sigmadust}
   \end{figure}

Fig.~\ref{fig:sigmadust} shows in addition to the gas surface density the total dust surface density and the surface density for the reference dust components shown in Fig.~\ref{fig:particles} and Fig.~\ref{fig:sigma}. The plot shows that the dust surface density is mainly determined by the larger particles which show efficient concentration at the pressure maximum. We also note that the density gap at 20 au is clearly induced by the outflow boundary as the surface density is truncated in a region between 20-25 au which causes a local pressure maximum and so the evolution of a vortex there. The actual inner dead-zone edge should be much closer inward.
\paragraph{Radiative transfer (RT):}
\label{sec:rt}
To calculate the temperature distribution and thermal re-emission maps
of the disks, we use the RT code MC3D
(\citealtads{1999A&A...349..839W,2003CoPhC.150...99W}). We focus on
the continuum emission of the dust phase of the disk assuming that the
dust material is composed of $62.5\, \%$ silicate and $37.5\, \%$
graphite (optical data by \citealtads{2001ApJ...548..296W}). We
calculate the wavelength-dependent absorption efficiencies for the
different grain sizes with Mie theory, using the tool MieX
\citepads{2004CoPhC.162..113W}. 

In this study the dust phase consists of two components. The first
component is the system of small dust particles that fully couple to
the motion of the gas in the disk (see Section~\ref{sec:hydro}). The
distribution of these particles follows the gas distribution. The second component of the dust phase is represented by
large dust particles (see Section \ref{sec:hydro}). As for the small
particles, the large particles initially follow the continuous power
law size distribution given by Eq.~(1). Each
grain of size $a_g$ in the simulation represents a number of
dust particles $\tilde{N}_g(a_g)$. Appendix~\ref{sec:ng} presents a detailed description of the calculation of
$\tilde{N}_g(a_g)$. 
\section{Results and Discussion}
\label{sec:results}

Our goal is to explore the impact of the large dust particles on the thermal re-emission of circumstellar disks. Our analysis is based on the simulated 
$441\, \rm \mu m$, $871\, \rm \mu m$, $1.3\, \rm mm$ and $2.0\, \rm
mm$ re-emission maps of our disks (assuming they are located at a
distance of $140\, \rm pc$ from the observer) at the two selected time steps. In particular, we compare these maps to our previous models with the small dust grains \citepads{Flock}.\par
\begin{figure}[!t]
	\centering
       \resizebox{\hsize}{!}{\includegraphics{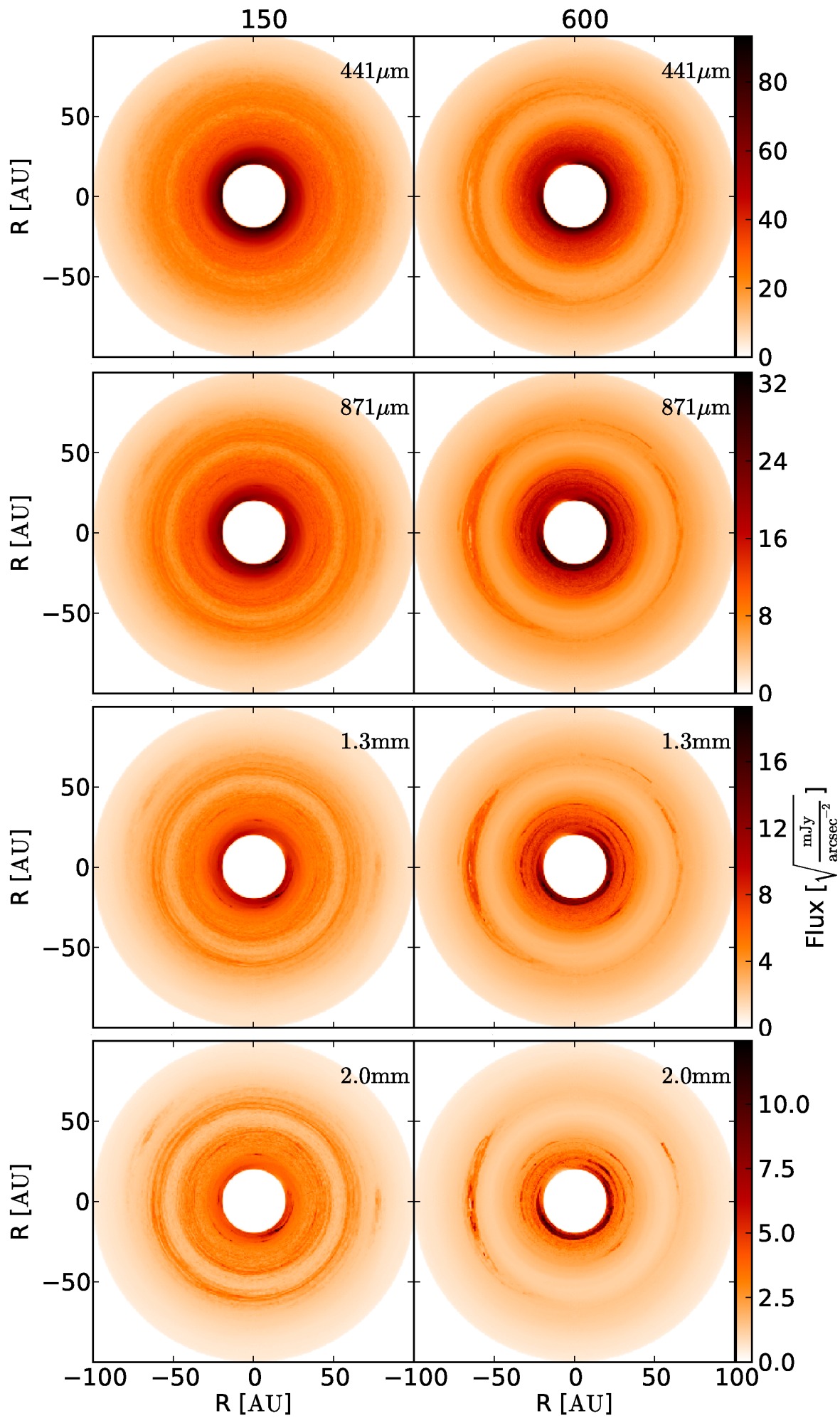}}
  \caption{Selected re-emission maps for the disk after $150$ inner orbits ($18,\!900$ years, left column) and $600$ ($75,\!600$ years, right column). Each wavelength ($441\, \rm \mu m$, $871\, \rm \mu m$, $1.3\, \rm mm$ and $2.0\, \rm mm$) is presented in its own line. The square root of thermal re-emission flux per arcsec$^2$ is shown color-coded. At both time steps and for every wavelength a gap is visible in the maps. Depending on time and wavelength the re-emission maps of the disk show different structures (rings and concentrations).}
  \label{fig:RTR}
   \end{figure}

The additional large dust grains increase the total flux of the thermal disk re-emission at $1.3\, \rm mm$ by a factor of $\approx 1.5$. 
This is due to the absorption efficiency of the large grains that is several orders of magnitude higher in comparison to that of the small grains at wavelengths $> 100\, \rm \mu m$\footnote{We note that the factor of 1.5 should be understood as a lower limit as we do not resupply the grains in the outer domain. Especially the particles (radii $\geq 3.1\, \rm mm$) drift quickly radially inward and leave the simulation.}. In scattered light at a wavelength of $2.2\, \rm \mu m$ (K band) the appearance of the disk is not influenced by the large dust particles at all, because they are all confined to less than one scale height (see Section \ref{sec:hydro} and Fig. \ref{fig:sigma}) and this part of the disk is not traced by scattered light at this wavelength \citepads{2013A&A...549A..97R}.\\
The large dust grains are not homogeneously distributed in the
disk. Their distribution is also time dependent (see Section  \,
\ref{sec:hydro} and Fig. \ref{fig:particles}). As a result, they are
able to change the (sub)mm appearance of a circumstellar disk
significantly within a few $10^4$ years (one inner orbit $\approx$ 126
years), even though they only contribute a fraction $\approx 0.07\%$ of the total mass of the disk (compare left and right column of Fig. \ref{fig:RTR}). This is in contrast with the results in Section 4.2 of \citetads{Flock}, where the re-emission structure of the disk -- with small grains only -- remains constant over the same period\par
\begin{figure}[t]
	\centering
       \resizebox{\hsize}{!}{\includegraphics{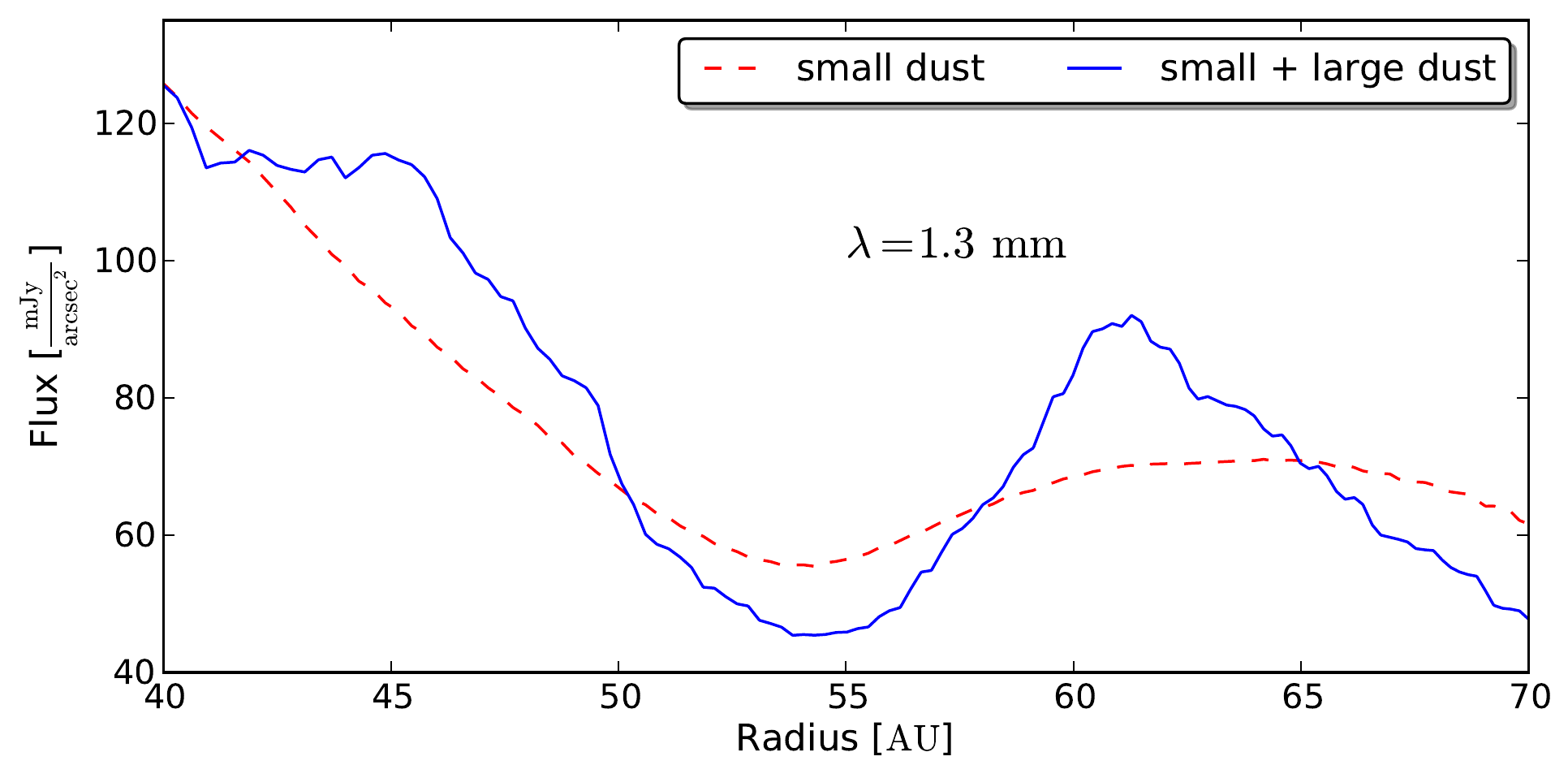}}
   \caption{Comparison of radial brightness profiles of the disk after
     $150$ inner orbits ($18,\!900$ years) with (solid blue line) and
     without additional large dust particles (dashed red line) at a
     wavelength of $1.3\, \rm mm$. The contrast ratio between the
     brightness minimum, which corresponds to the gap center, and the
     brightness maximum of the disk for radii $> 55\, \rm au$ is
     enlarged by a factor of $\approx 1.5$ through the additional
     large dust particles in comparison to the case with small grains
     only. The total number of grains particles $N_{max}$ is the same
     for both the blue and the red curve (see Appendix~A), which means
     that the number of small grains used to compute the red curve is
     reduced in comparison to the blue lined case. This leads to a
     decreased flux at the gap center.}
   \label{fig:profile}
   \end{figure}

 
The thermal emission maps in Fig. \ref{fig:RTR} show three dominating structures, which are due to the distribution of the (large) dust in the disk (see Fig. \ref{fig:particles}). At both time steps, the gap, which was discussed already by 
\citetads{Flock} is present in the re-emission maps. After $150$ inner orbits ($18,\!900$ years) the large dust particles have accumulated into rings. In the thermal emission maps this leads to an enlarged brightness contrast ratio between the innermost region of the gap and its edges. In comparison to the results obtained without the larger dust grains the contrast is $\approx 1.5$ times higher (see Fig. \ref{fig:profile}). The gap appears deeper in this case. Similar results have been found for the local concentrations of larger, weakly coupled dust grains in the case of planet-disk interaction by \citetads{2004A&A...425L...9P,ric06,fou07,lyr09,fou10,ayl12,zhu12}. \par
After $600$ inner orbits ($75,\!600$ years) the disk shows a large
vortex, which becomes visible due to the concentration of larger
particles (see Fig. \ref{fig:particles}). With the exception of the
$50\, \rm \mu m$ particles, the vortex has destroyed the dust ring
structures and concentrated the larger dust
particles azimuthally. This size-dependent dynamics of the dust grains lead to
different thermal emission maps at different wavelengths (see
Fig. \ref{fig:RTR}, right). Observations at longer wavelengths ($\geq
1.3\, \rm mm$) trace the dust concentration in the vortex very well,
while the radiation of the smallest particles in our setup dominates
at shorter wavelengths and is more smoothly distributed (see
Fig. \ref{fig:RTR}, right). A similar trend was seen in a recent observations of HL Tau at $7\, \rm mm$ \citep{car16}, which shows clear azimuthal variations inside the ring at $7\, \rm mm$ compared to the smooth emission at $1.3\, \rm mm$. Eventually, this offers the possibility to
determine the size of the particles in the vortex by using spatially 
resolved maps of the spectral index (see Section
\ref{sec:discussion}).\par 
As already mentioned in Section \ref{sec:rt}, the set of large dust particles samples a continuous grain size distribution. 
Consequently, the distribution of dust grains of a certain size, that
is not included in our setup, can be approximated by interpolating
from the upper and lower size bins. We have checked that a higher
sampling of the dust size bins has no effect onto the disk structures. 

We note again that the vortex lifetime is around 40 local orbits until it is destroyed and reformed again on a timescale of 20 local orbits \citep{Flock}. 
We do not observe any migration of the vortex. We point out that we consider Ohmic resistivity as the sole source of magnetic field dissipation and neglect for simplicity the Hall effect and ambipolar diffusion. Recent results have shown their importance for MRI--driven turbulence in circumstellar disks \citep{bai13,les14,bet16}. They could potentially shift the MRI inactive zone radially outward or even trigger new instabilities. In addition, in this region the electron heating could become important, which was recently shown by \citet{mor15}.\par  

\section{Comparison to recent ALMA observations}
\label{sec:discussion}

\begin{figure*}[htbp]
\mbox{
   \includegraphics[width=8.6cm]{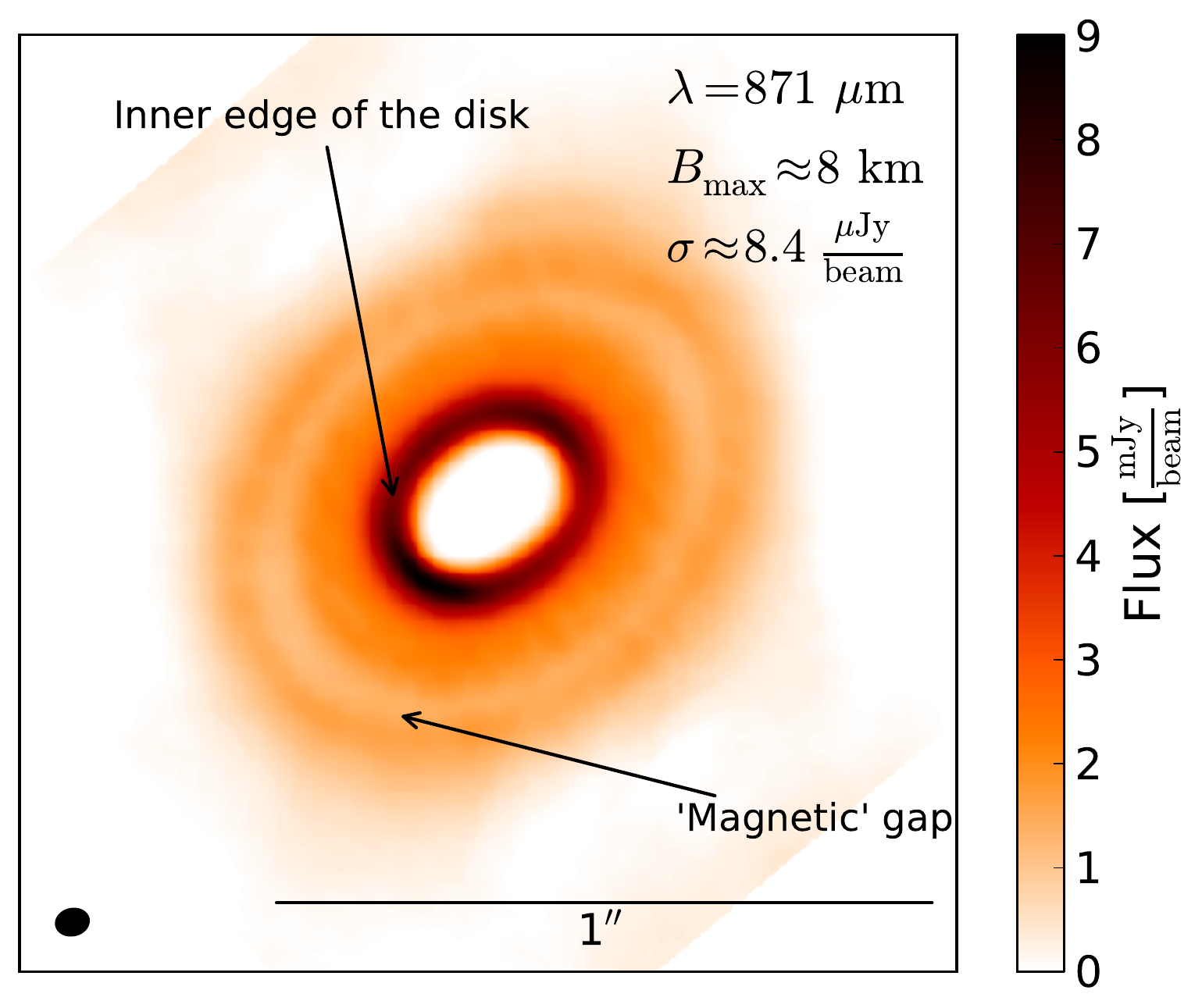}
   \includegraphics[width=9cm]{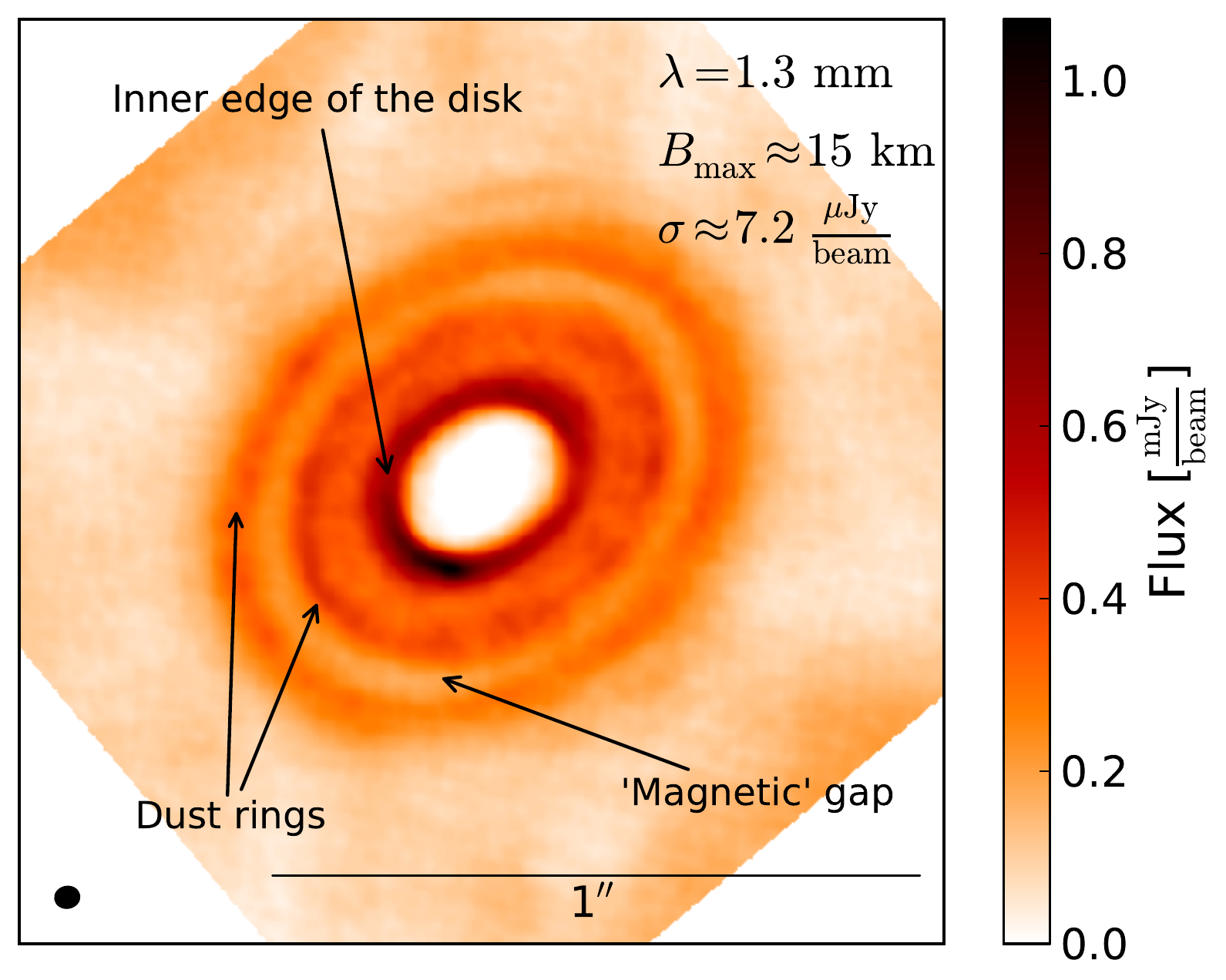}}\\
\mbox{
   \includegraphics[width=9cm]{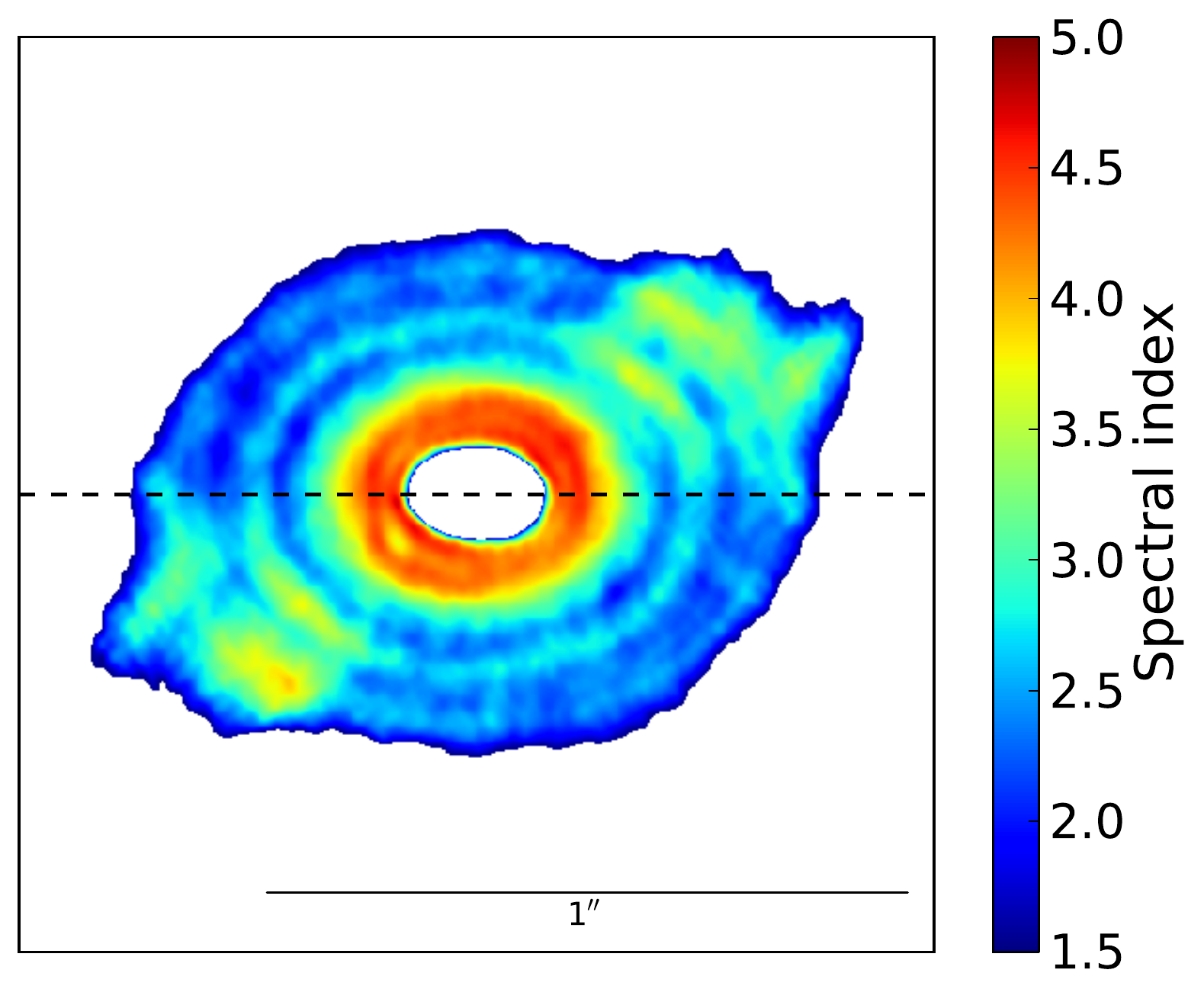}
   \includegraphics[width=8cm]{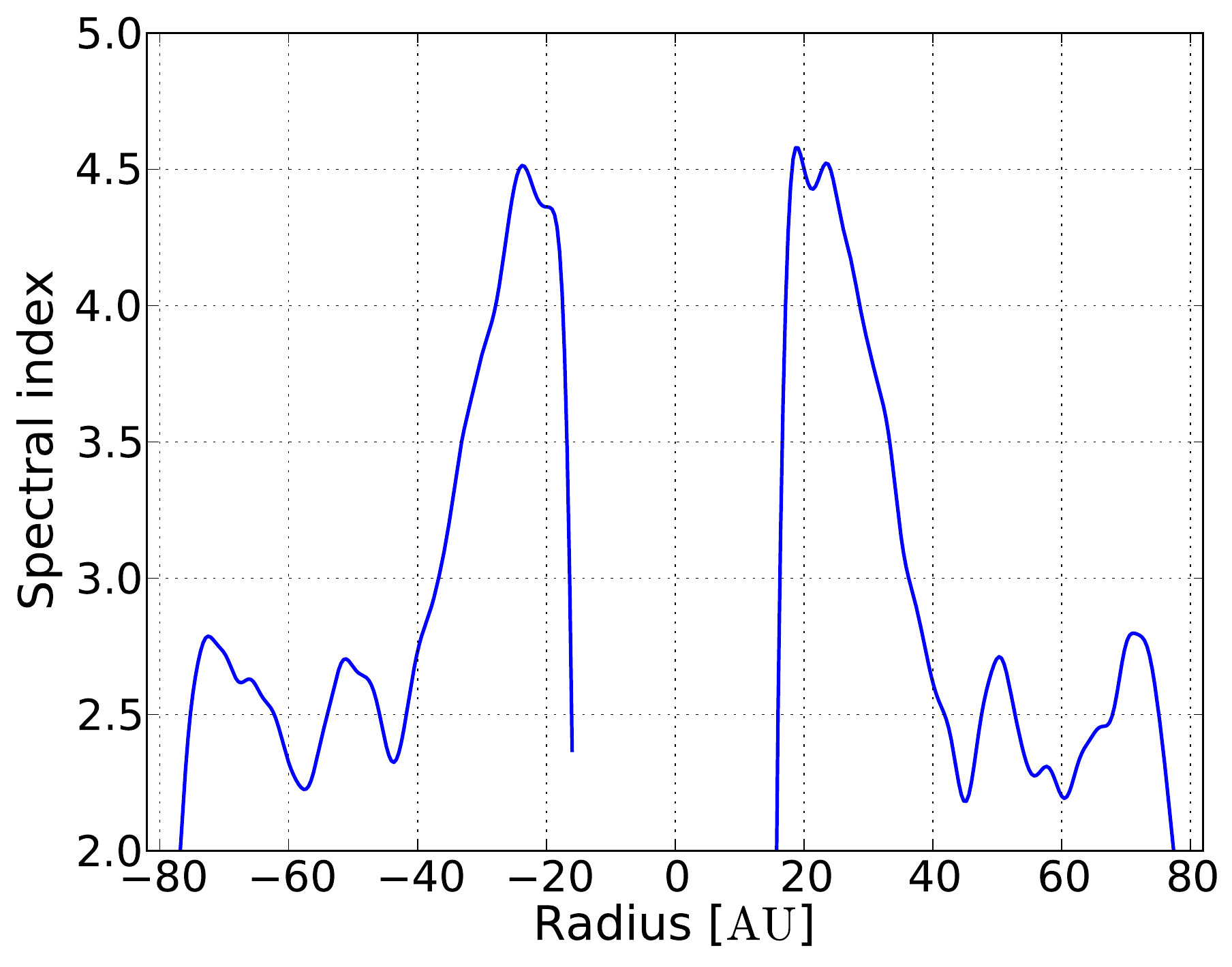}}
\caption{Top: Simulated ALMA images of the disk after $150$ inner
  orbits ($18,\!900$ years) at a wavelength of $871\, \rm \mu m$
  (left) and $1.3\, \rm \mu m$ (right). The position, the disk
  inclination, distance and stellar luminosity of HL Tau were used for
  these simulation. At both wavelengths a gap is visible. The gap in the images are
  emphasized and we note that its origin lies in the magnetic field in the disk and not
  a planet. Bottom
  left: Spatially resolved map of the spectral index calculated from
  the simulated observation on the top. The map is rotated clockwise
  by $40^\circ$ in comparison to the maps at the top. Bottom right:
  Profile of the spectral index along the cut indicated by a dashed
  line on the bottom left figure.}
\label{fig:ALMA}
\end{figure*}

\paragraph{Dust ring structures:}
The ring structures found in the density distribution of the disk after $150$ inner orbits ($18,\!900$ years) appear qualitatively similar to the structures recently observed in the disk of HL Tau \citepads{2015arXiv150302649P}. Therefore, we explore how an equivalent (simulated) ALMA observation of our disk model would look like. We select the position of HL Tau (RA: $4\, \textrm{h}31\textrm{min}38.4\textrm{s}$,  DEC: $+18^\circ 13\arcmin57.4\arcsec$ J2000, distance: $140\, \rm pc$; \citealtads{2011ApJ...741....3K}) and its inclination of $40^\circ$ \citepads{2011ApJ...741....3K}. The thermal emission maps are re-calculated under the assumption of the stellar luminosity ($L_\star \approx 8.3\, \rm L_\odot$) and an effective temperature ($4000\, \rm K$) of HL Tau \citepads{2011ApJ...741....3K}. We focus on the wavelengths of $871\, \rm \mu m$ and $1.3\, \rm mm$, and select very extended configurations of the ALMA array with maximum baselines of $\sim 8\, \rm km$ and $\sim 15\, \rm km$. The observing time is chosen to be $6\, \rm h$. The resulting spatial resolution is $49\, \textrm{mas} \times 39\, \textrm{mas}$ at $871\, \rm \mu m$ and $37\, \textrm{mas} \times 32\, \textrm{mas}$ at $1.3\, \rm mm$. Thermal and phase noise are added (see the details on our ALMA setup in \citetads{Flock}. 
We emphasize that the goal of these simulations is to study
qualitatively the extent to which the ring structure found in our simulations resembles the structure found in the HL Tau disk, while a quantitative fit to the observational results of HL Tau which would require a detailed consideration of the envelope which has been investigated in RT models by \citetads{1999ApJ...519..257M}.\par
ALMA enables the detection of single dust rings with this
configuration for an observing wavelength of $1.3\, \rm mm$ (see Fig
\ref{fig:ALMA}, top right). At a wavelength of $871\, \rm \mu m$ the
dust ring structure is less pronounced (see Fig \ref{fig:ALMA}, top left). This is due to the emission of the smaller grains that are
distributed homogeneously in the disk and still efficient at this
wavelength (Fig. \ref{fig:RTR}). These differences offer a possibility
to identify dust concentrations as the origin of the ring
structures. From the wavelength-dependence of the emission, one can constrain the size of the re-emitting particles. For this purpose, we calculate the spatially resolved map of the spectral index from both simulated ALMA observations (see Fig \ref{fig:ALMA}, bottom left) and the profile of the spectral index in a selected cut through the map (see Fig \ref{fig:ALMA}, bottom right). The spectral index is defined as:
\begin{equation}
 \alpha_\text{Sp} = - \cfrac{\partial \log_{10}\left(F_\lambda\right)}{\partial \log_{10}\left(\lambda\right)}.
\label{glg:spektralindex}
\end{equation}
At the position of the dust rings the spectral index is reduced, which indicates the presence of larger particles. Within the gap in the disk, the spectral index is increased which indicates that only smaller dust grains are present in this region. 
We note that the asymmetric feature, seen in the spectral index map could arise from the projection as the disk is inclined along that direction.
Finally, we conclude that similar (sub)mm appearances of circumstellar
disks can result from an inhomogeneous grain size distribution in the
radial direction as discussed above without the necessity to invoke
the presence of a planet as is usually done. 
We emphasize that the position of the gap and ring structure is located at the dead-zone outer edge, which depends mainly on the surface density amount and profile 
of gas and dust \citep{dzy13,Flock}. For the parameters used here we expect this to happen at any location where the total surface density becomes around 15 g cm$^{-2}$, assuming a dust to gas mass ratio of 0.01, as this value marks the threshold for MRI activity. We note again that the ring state, presented in Fig. \ref{fig:ALMA}, is present more often than only once in
the simulation. As we have pointed out, the lifetime of the vortex is around 40 local orbits. After this
time, the particles will be again only radially concentrated. It is true that each individual ring state
will look different, especially if we do not resupply the particles inside the domain. The vortex state is still more likely to observe (2/3 vortex state vs. 1/3 ring state).

\paragraph{Dust concentrations in vortices:}

   \begin{figure}[t]
	\centering
       \resizebox{\hsize}{!}{\includegraphics{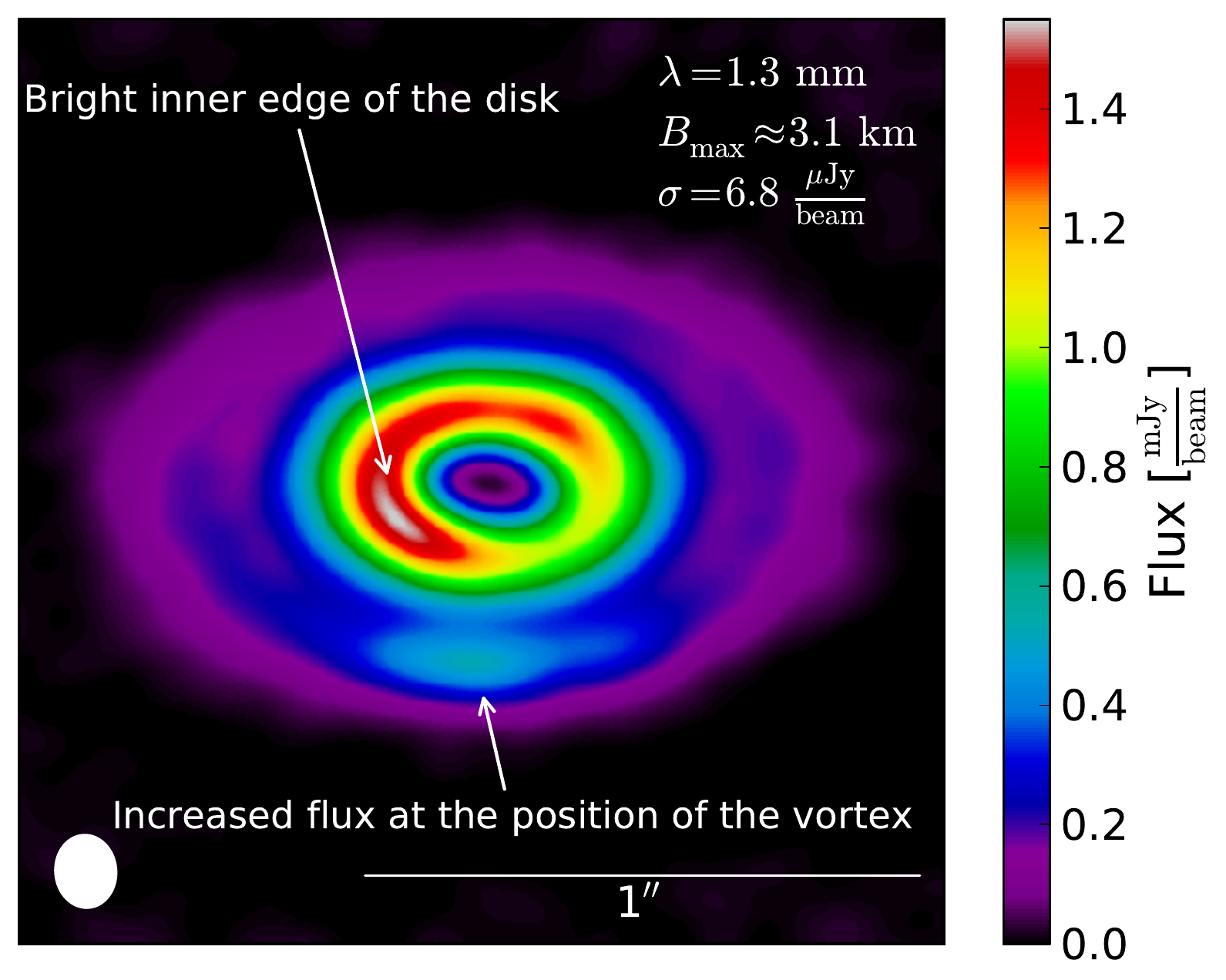}}
   \caption{Simulated ALMA images of the disk after $600$ inner orbits ($75,\!600$ years) at a wavelength of $1.3\, \rm mm$. The object's position, the disk inclination and distance of Oph IRS 48 have been used. In comparison to Fig. \ref{fig:RTR}, the model was rotated by $90^\circ$ counterclockwise. We note that the inner edge of the disk represents the beginning of the computational domain. } 
   \label{fig:IRS48}
   \end{figure}

Due to the concentration of large dust particles in the vortex, the re-emission maps of the disk after $600$ inner orbits ($75,\!600$ years) appear similar to various recent ALMA observations of young circumstellar disks (see section~\ref{sec:intro}). In order to investigate this similarity, we explore how a simulated ALMA observation of this disk model will look like. We select the position, distance and object inclination of Oph IRS 48 (RA: $16\, \textrm{h}27\textrm{min}37.8\textrm{s}$, DEC: $-24^\circ 30\arcmin35.3\arcsec$ J2000, distance: $120\, \rm pc$, inclination: $50^\circ$; \citealtads{2013Sci...340.1199V}). Other model parameters such as stellar temperature and luminosity or disk mass are not adapted to the parameters of Oph IRS 48.
Our goal is to show the feasibility to observe the vortex in the disk. In comparison to the thermal emission maps in the right column of Fig. \ref{fig:RTR}, we rotate the model counterclockwise by $90^\circ$. We select a wavelength of $1.3\, \rm mm$, an observing time of $5\, \rm h$, and a configuration of the ALMA array with a maximum baseline of $3.1\, \rm km$, corresponding to a spatial resolution of $0.13\, \rm \arcsec \times 0.11\, \rm \arcsec$.\par
The concentration of large dust particles in the vortex increases the
flux at the vortex position by a factor $\approx 2.8$ in the simulated
ALMA observation compared to the models without large dust grains. The peak flux at the vortex compared to the background is around 10, which is lower then the value 130 found by \citep{zhu14}. However we note that several factors influences this peak emission, especially the dust concentration and size of the vortex. E.g. the planet induced vortex by \citep{zhu14} appears to be larger in size.
Finally, we show the feasibility to detect a
vortex through its ability to concentrate large dust grains in a
circumstellar disk again, without the necessity of a planetary mass
perturber. 

\section{Conclusion}
\label{sec:conclusions}
In this work, we have followed the motion and concentration of large
dust particles with different sizes in a typical protoplanetary disk
using a non-ideal 3D MHD disk simulation.
We have performed RT calculations to derive the temperature profile and thermal emission maps of the disk. Our goal was to investigate the impact of (sub)mm-sized dust grains on spatially resolved thermal emission maps of circumstellar disks and the feasibility to detect selected structures with ALMA. In summary, we find:
\begin{itemize}
 \item The larger dust particles (with radii ranging from $50\, \rm \mu m$ to $1\, \rm cm$) are weakly coupled to the gas motion and accumulate in rings.
 \item Vortices in the disk can break up those rings and concentrate the dust particles azimuthally. This leads to an increased local dust-to-gas ratio in the vortex center.
 \item Due to the high efficiency of emission of these large dust
   grains, the dust rings as well as the dust concentration in
   vortices are emphasized in the thermal emission maps of the
   disk. For example, the higher concentration of the larger dust grains enlarges the brightness contrast between the minimum in the gap and the peak outside by at least a factor of 1.5 at a wavelength of $1.3\, \rm mm$ in comparison to the results of \citetads{Flock}.
 \item Depending on the wavelength, the gap, dust rings and dust concentration in vortices can be observed with ALMA.
 \item The spatial distribution of dust grains of different size can also be identified by a decreased spectral index in spatially resolved spectral index maps calculated on the basis of simulated ALMA observation.
 \item The ring structures as well as the dust concentration in the
   vortex are stable for at least $10^4$ years. In addition, the disk
   is alternating between a state with and without azimuthal dust
   concentration by a vortex, as already shown by \citetads{Flock},
   significantly increasing the likelihood to observe the presented
   structures. The vortex state is more likely to observe (by 66\%) compared to the ring state.
\end{itemize}
We conclude that inhomogeneous and non-axisymmetric distributions of
larger dust grains have a major impact on the (sub)mm appearance of a
circumstellar disk. Our work shows that such inhomogeneities can
naturally arise as a result of internal dynamical processes associated
with the presence of a weak magnetic field, without having to rely on
a planet orbiting in the disk. More constraints will be required in
the future to disentangle the origin of the observed (sub)mm
structures in protoplanetary disks, including observations of the
magnetic field \citepads[e.g., HL Tau][]{2014Natur.514..597S} and/or
of the disk kinematics \citep{sim15}.

\begin{acknowledgements}
We thank the anonymous referee for providing extensive and constructive comments, which significantly improved the paper. We acknowledge financial support by the German Research Foundation (J.P. Ruge: WO 857/10-1; H.H. Klahr: KL 14699-1). Mario Flock, Sebastien Fromang, and Heloise Meheut are supported by the European Research Council under the European Union’s Seventh Framework Programme (FP7/2007-2013) / ERC Grant agreement nr. 258729. Parallel computations have been performed on the IRFU COAST cluster located at CEA IRFU. \end{acknowledgements}
\bibliography{bib2}
\appendix
\section{Calculating $\tilde{N}_g$}
\label{sec:ng}
In the case of the large dust grains, each of the initial 'test' particles of a certain radius $a_g$ represents a certain number of dust particles $\tilde{N}_g(a_g)$.
We assume, that the initial dust mass $M_\text{d}$ amounts to $1\%$ of the total gas mass $M_\text{gas}$ and is the sum of the mass of the small dust $M_\text{sd}$ and the mass of the larger dust $M_\text{ld}$.
In the radiative transfer calculation, the dust properties of the small dust are discretized into $1000$ bin sizes, therefore the mass $M_\text{sd}$ can be expressed in the following way:
\begin{equation}
  M_\text{sd} = \cfrac{4}{3}\, \pi \, \rho_\text{md} \sum_{i=1}^{1000} a_i^3 \cdot N(a_i),
\end{equation}
where $\rho_\text{md} = 2.7\, \rm g.cm^{-3}$ is the density of the dust material and $N(a_i)$ is the absolute number of particles with radius $a_i$. Following the grain size distribution (see Eq. \ref{eq:grainsizedist}), $N(a_i)$ can be written as a fraction of a reference number of particles. Without loss of generality, we select the total amount of particles $N_\text{max}$ with radius $a_\text{max}$:
\begin{equation}
 N(a_i) = N_\text{max} \cdot \left(\cfrac{a_i}{a_\text{max}}\right)^{-2.5}\label{glg:nai}.
\end{equation}
The mass of the large dust grains can be described in the same way:
\begin{equation}
  M_\text{ld} = \cfrac{4}{3}\, \pi \, \rho_\text{md} \sum_{g=1}^{10} a_g^3 \cdot N(a_g),
\end{equation}
Because of Eq. \ref{glg:nai}, we finally get:
\begin{equation}
0.01\, M_\text{gas} = M_\text{d} = M_\text{sd} + M_\text{ld} = \cfrac{4}{3}\, \pi \, \rho_\text{md} \cfrac{N_\text{max}}{a_\text{max}} \left(\sum_{i=1}^{1000} a_i^{0.5} + \sum_{g=1}^{10} a_g^{0.5}\right)\label{glg:staubmasse_flock}.
\end{equation}
Through reordering of Eq. \ref{glg:staubmasse_flock}, $N_\text{max}$ -- and therefore every $N(a_i)$ and $N(a_g)$ -- can be calculated. Finally, $\tilde{N}_g$ is given by:
\begin{equation}
 \tilde{N}_g = \cfrac{N(a_g)}{50\,000}.
\end{equation}
To calculate the dust surface density one simply has to integrate over a radial bin size $\Delta r$ and then calculate the given dust mass per unit area: \begin{equation} 
\Sigma_{dust}= \frac{\int_{\Delta r} M_\text{d}\, dr} {\int_{\Delta r} \int_\phi r\, dr\, d\phi}.
\label{glg:surface_density}
\end{equation}

\section{Particle method validation}
\label{a:benchmark}

\begin{figure}
  \resizebox{\hsize}{!}{\includegraphics{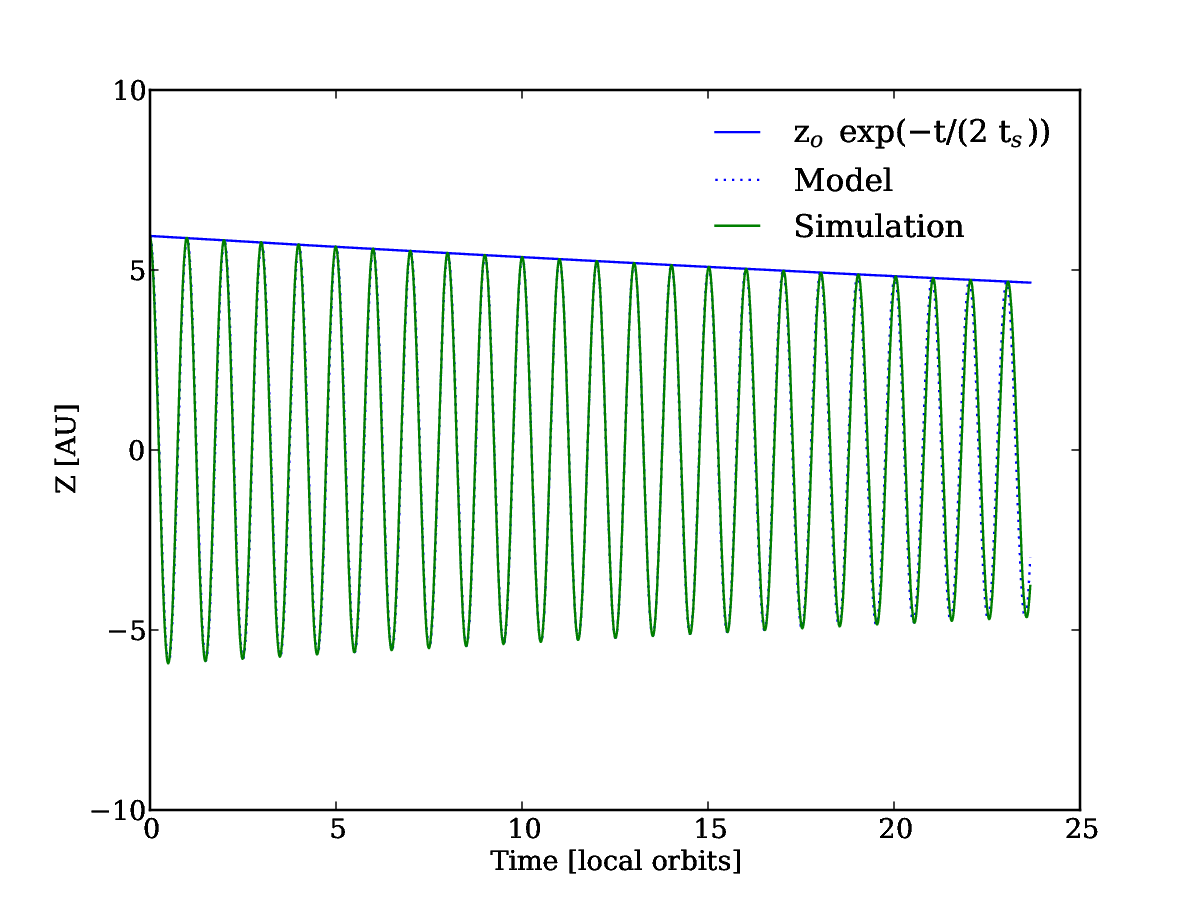}}
\caption{Vertical motion of a particle with Stokes number of $\rm St=300$ (green solid line), a model of a damped oscillator with orbital frequency period (dotted line) and the amplitude damping $z(t)$ (blue solid line) plotted over time.}
\label{fig:vert_set}
\end{figure}

In this section, we performed global hydrodynamical simulations to validate our method. For the test models, we use the setup by \citet{nel13} which describes the hydrostatic disk solution dependent on the two parameter p and q. 

\begin{eqnarray}
\rho&=& \rho_0 \left ( \frac{R}{R_0} \right )^p \cdot e^{\frac{G M}{c_s^2} \left[ \frac{1}{r} - \frac{1}{R} \right ]}\\
 v_\phi&=& R \Omega_K \cdot \left ( 1.0+(p+q) \cdot \frac{H}{R}^2 + q \cdot (1.0-\sin \theta) \right )^{0.5},
\end{eqnarray}
with the cylindrical radius $R=r\sin(\theta)$, the spherical radius $r$, the orbital frequency $\Omega_K$, the sound speed $c_s=H \cdot \Omega$ and the scale height $ H=H_0 (\frac{R}{R_0})^{(q+3)/2}$. For both tests, we use a global disk model with a radial extent of $20-40\, au$, $\Delta \theta=0.72$ radian and $\Delta \phi=0.4$ radian. The resolution is set to $(96x96x48)$ in $(r,\theta,\phi)$ and $H_0=0.1$ at $R_0=40\, au$. 
In the following two testruns, we check the vertical and radial integration and drag regimes individually.
\subsection{Vertical motion}
The first disk model uses the parameter set of $p=q=0$ so that the radial pressure gradient vanishes. In this case, the particles do not drift and we can focus on the vertical motion. In this test, we inject a large particle (Stokes number of 300) at two scale heights, located at 30 au. The vertical motion of large particles can be described by a damped oscillation, with oscillation frequency $\omega=\Omega_K$ and the amplitude damping $z(t)$ 
\begin{equation}
z(t)=z_0\cdot e^{-t/(2\tau_t)},
\end{equation}
with the the stopping time $\tau_t$. Fig.~\ref{fig:vert_set} shows the particle evolution over height. The particle oscillates with orbital rotation frequency and its motion fits very well a damped oscillation.
\begin{figure}
  \resizebox{\hsize}{!}{\includegraphics{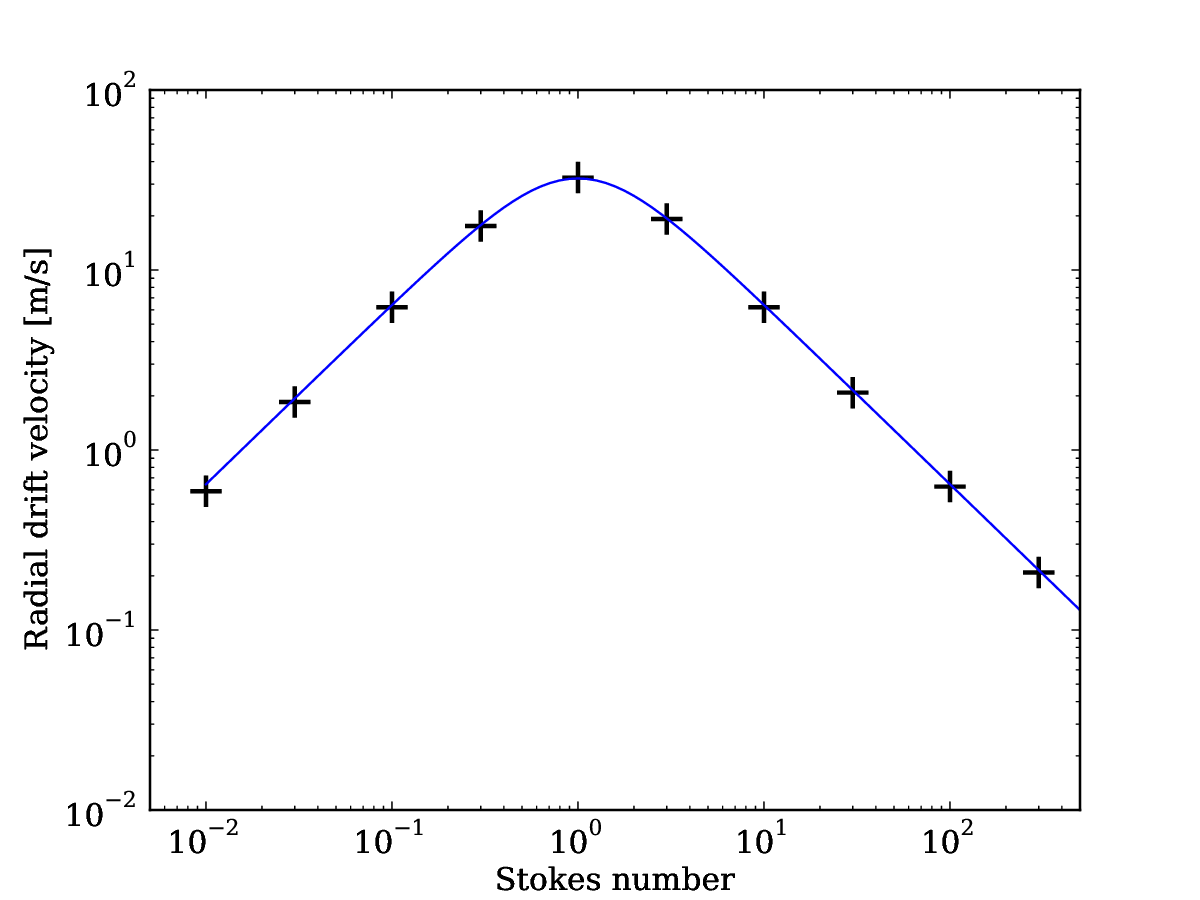}}
\caption{Averaged radial particle velocity (black crosses) with Eq.~\ref{eq:vdust} overplotted (blue solid line).}
\label{fig:rad_vel}
\end{figure}
\subsection{Radial motion}
In the second test we inject particles of different sizes at the midplane at $R=35 au$ to study the radial drift. For this setup we use $p=-1.0$ and $q=-0.5$. We set the particles initial velocity to Keplerian. The particle drift is plotted in Fig.~\ref{fig:rad_vel}. The particles with Stokes number $\rm St=1$ drift radially inwards. The maximum drift velocity $v_{max}$ is proportional to the velocity difference $v_{K}-v_{\phi}$ and is in our case around 33 meters per second. The drift velocity of different particles sizes can be expressed using
\begin{equation}
v_{dust}=\frac{-2 v_{max}}{\mathrm{St}+\frac{1}{\mathrm{St}}}.
\label{eq:vdust}
\end{equation}
We average the particle velocity between 2 and 8 local orbits. Fig.~\ref{fig:rad_vel} shows the particle velocity, overplotted with Eq.~\ref{eq:vdust}. The velocities match very well the prediction of Eq.~\ref{eq:vdust}. 

\subsection{Benchmark: The radiative transfer}
\label{ap:benchrt}
\begin{figure}
     \vspace{-1.5cm}
     \resizebox{0.9\hsize}{!}{\includegraphics{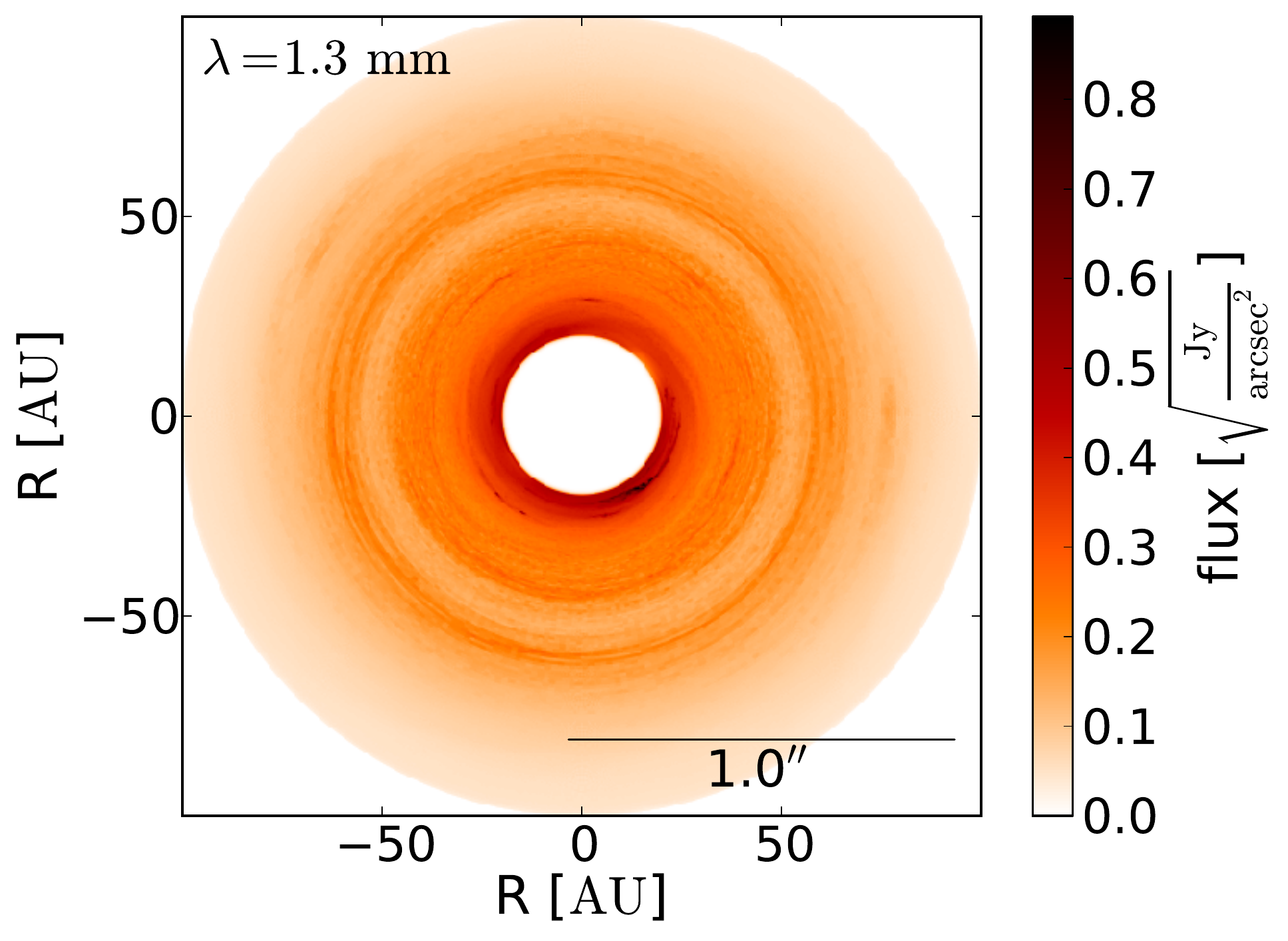}}
       \resizebox{0.9\hsize}{!}{\includegraphics{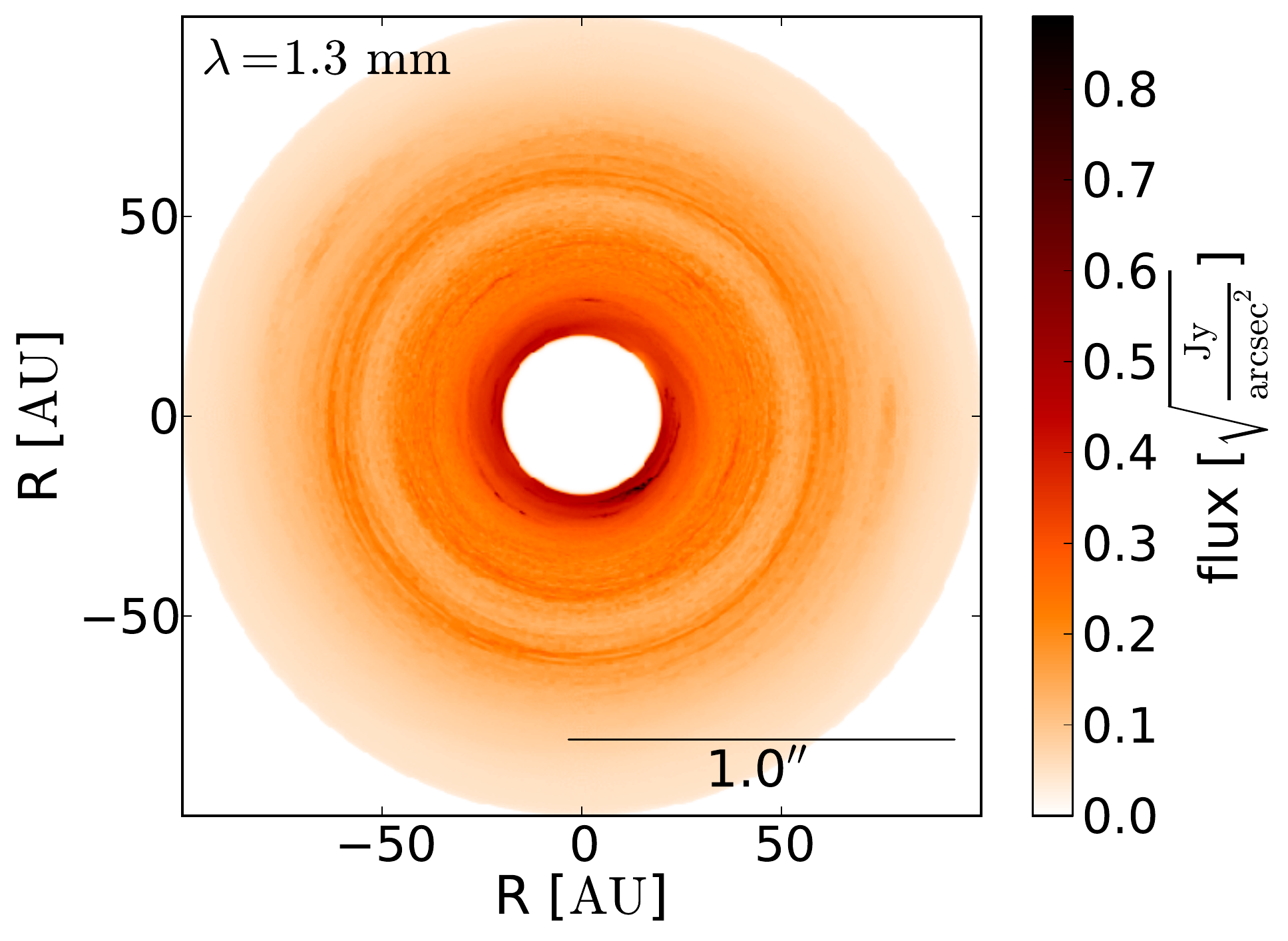}}
       \resizebox{0.9\hsize}{!}{\includegraphics{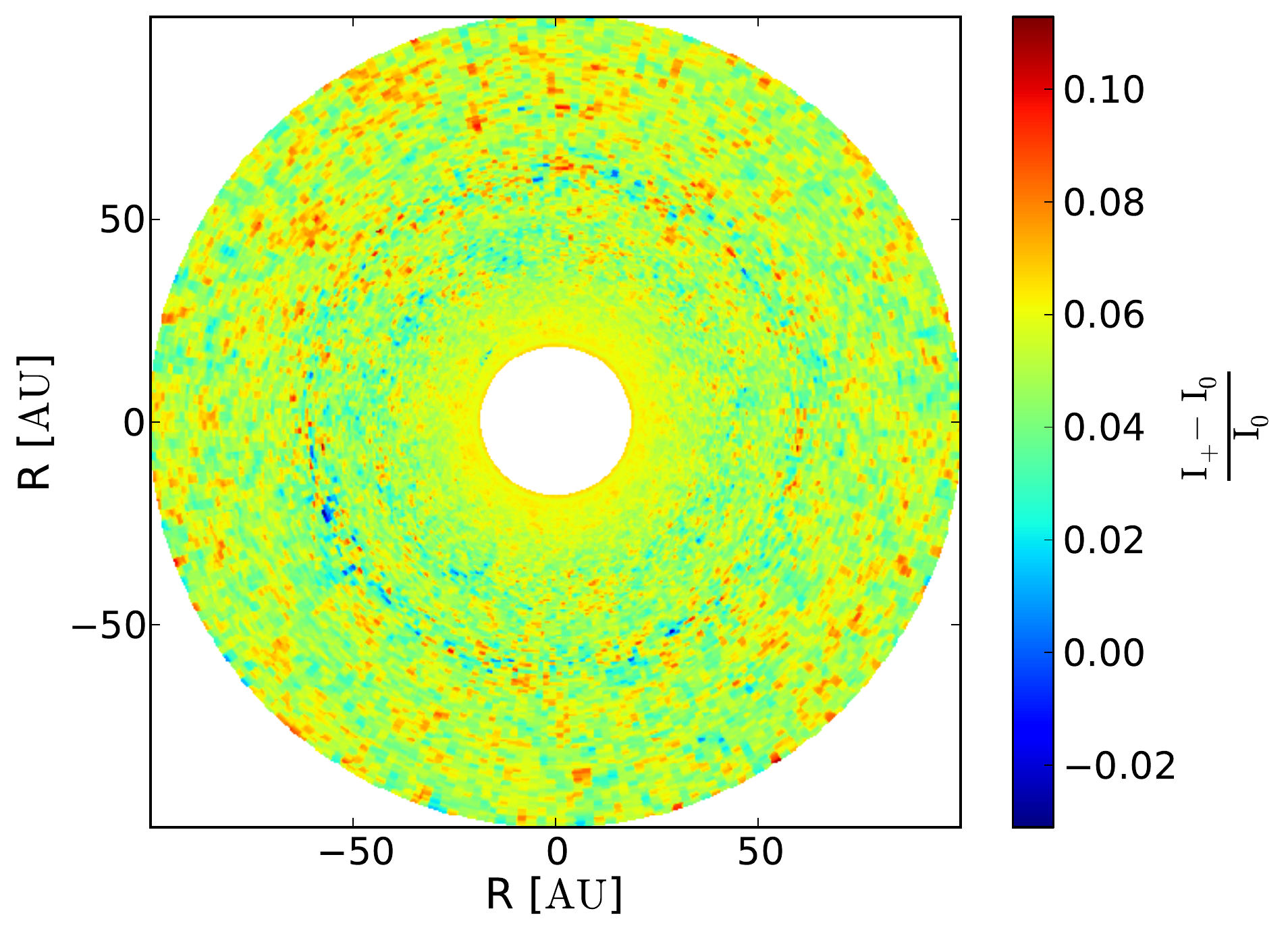}}
  \caption{Selected re-emission map at a wavelength of $1.3\, \rm mm$ of a disk model including $5 \cdot 10^5$ (top), $5 \cdot 10^6$ (middle) initial larger dust particles and the relative intensity deviations between those two models (bottom). The differences show deviations in the order of several percent.}
  \label{fig:REV-1300um}
   \end{figure}

 Following the MHD simulation presented in Appendix \ref{sec:hydro}, we performed an additional global simulation with $5 \cdot 10^6$ particles. For the setup, we use the identical same initial conditions for the MHD setup as the previous model. To distribute the larger particles we use a $1/R$ profile instead of a uniform distribution. We use the concept, which has been already defined in Section \ref{sec:ng}, to incorporate these particles into the radiative transfer simulation. All further parameters remain constant. The gas evolution and so the distribution of the small particles is the same as as in the previous models. The larger particles quickly redistribute and are concentrated at the same position. We perform the radiation transfer calculation at the same time output after 150 inner orbits.\\ 
The results of both radiative transfer calculations at a wavelength of $1.3\, \rm mm$ are presented in Fig.~\ref{fig:REV-1300um}. Both simulations show a very similar emission map with differences in the order of several percent, see Fig.~\ref{fig:REV-1300um} bottom. From that we conclude, that the influence of the number of large particles and the initial distribution remains small. This comparison strengthens our conclusion as the structures resulting from the positions of the larger dust particles (see section \ref{sec:results}) are robust. 

\end{document}